# Introducing Morphit, a new type of spreadsheet technology

Ted Hawkins Ph.D., Andrew Lemon Ph.D., Alec Gibson B.Sc. (Hons) (Dunelm)
The Edge Software Consultancy Ltd.,
77 Walnut Tree Close, Guildford, Surrey. GU1 4UH UK
thawkins@edge-ka.com

## ABSTRACT

*This paper describes a new type of spreadsheet which mitigates the errors caused by incorrect range referencing in formulae. This spreadsheet is composed of structured worksheets called tables which contain a hierarchical organization of fields. Formulae are defined at the field-level removing the need for positional references. In addition, relationships can be defined between fields in tables, allowing data to be modeled rather than simply processed and providing a re-usable framework for authoring spreadsheets. We shall describe the key features of tables with an emphasis on error detection and avoidance.*

## 1. INTRODUCTION

Despite their utility it is clear that in many cases current spreadsheets are not fit for purpose in today's business environment [Saadat 2008]. Inadvertent errors are too easily made [Chadwick 2008a, Dunn 2010, Panko 2008a], and auditing spreadsheets is so difficult that they have been identified as a significant factor in the financial crash of 2008 [Croll 2009] as well as a number of fraud cases [Butler 2009, Mittermeir et al. 2008, Panko and Ordway 2008, Saadat 2008]. Cell error rates have been estimated at between 2% and 5%, with approximately 94% of spreadsheets containing errors [Panko 2008a, Stephen G. Powell, Kenneth R. Baker 2009].

The reliance on explicit positional references in formulae is the root cause of two major sources of error - physical area related errors [Ayalew and Mittermeir 2008] and semantic and extendibility errors caused by poor layout [Przasnyski et al. 2011]. Consider the simple formula:

= SUM(A1:B2;A12:B13)

This formula is intrinsically difficult to read and understand which data the ranges are referring to, even more so were the references to go across worksheets. Positional formulae are often fragile with respect to worksheet modifications, errors can easily be introduced as new cells are inserted between the cells in a range, or formulae are incorrectly over-written as a result of copying absolute cell references. Auditing spreadsheets is a difficult and time consuming process since cell-by cell inspection is the only guaranteed method of ensuring compliance [Panko 2008b]. While we acknowledge that there are tools available to assist with the auditing process, it is our experience from the pharmaceutical industry, that very few people use them. Rather than trying to detect errors every time a spreadsheet is modified, we believe it is better to address the underlying flaws in the spreadsheet concept that allow these errors to be introduced in the first place.

Many of the problems with spreadsheets arise from the attempt to represent structured data in what is effectively an unstructured environment. Any structure to the data in a spreadsheet is implicit from a combination of user defined layout and presentation, rather than being inherent in the design.

For these reasons we have designed Morphit [Edge 2013] (www.edge-ka.com/products/morphit), a standalone application that, while retaining many of the features of traditional spreadsheets, incorporates the idea of explicitly defined data structure. In this paper we shall illustrate how this works in practice.

## 2. TABLES

Morphit contains two principal page types – *sheets* and *tables*. *Sheets* are almost identical to a traditional spreadsheet and used for presentation of static reports, however most work is achieved using *tables*. The quickest way to understand *tables* is to look at an example.

The screenshot above shows a Morphit *table* containing sales data for a pet shop. The *table* is defined as a hierarchy, with the top level relating to years, the second level relating to months, and the bottom level containing sales by code. New rows can be added at any level to expand the years, months, and sales codes.

Columns in a *table* are called *fields*. Of particular interest in this *table* are the two fields 'Monthly Total' and 'Yearly Total'. Both of these *fields* are calculated using exactly the same formula – '=SUM(Total)'. Morphit will calculate the result of this formula using only the values of 'Total' which are beneath the formula-containing cell in the hierarchy.

This already helps remove some of the common causes of errors in spreadsheets. The formula is only entered once for the 'Monthly Total' and 'Yearly Total' *fields* and it is propagated to any new cells in those *fields*. This removes the requirement to enter formulae in each cell, so eliminating many physical area related errors caused by transcription [Saadat 2008]. In addition, should any new rows be added at the Sales level, they will automatically be included in the calculations. This helps overcome the 'physical area mix up problem' [Ayalew and Mittermeir 2008],

The overall effect of this is that while the formulae should still initially be validated, the number of formulae to check has been reduced to an achievable level. In addition there is no longer any need to revalidate the *table* when adding new rows.

**Using Multiple Tables**

It is rare to find a problem simple enough that it can be modeled using a single *table*, or even a set of unconnected *tables*. Morphit contains powerful techniques that allow you to join several *tables* together in order to model complex business problems.

**Field Borrowing**

The first technique is called *Borrowing*. This is illustrated using the following example.

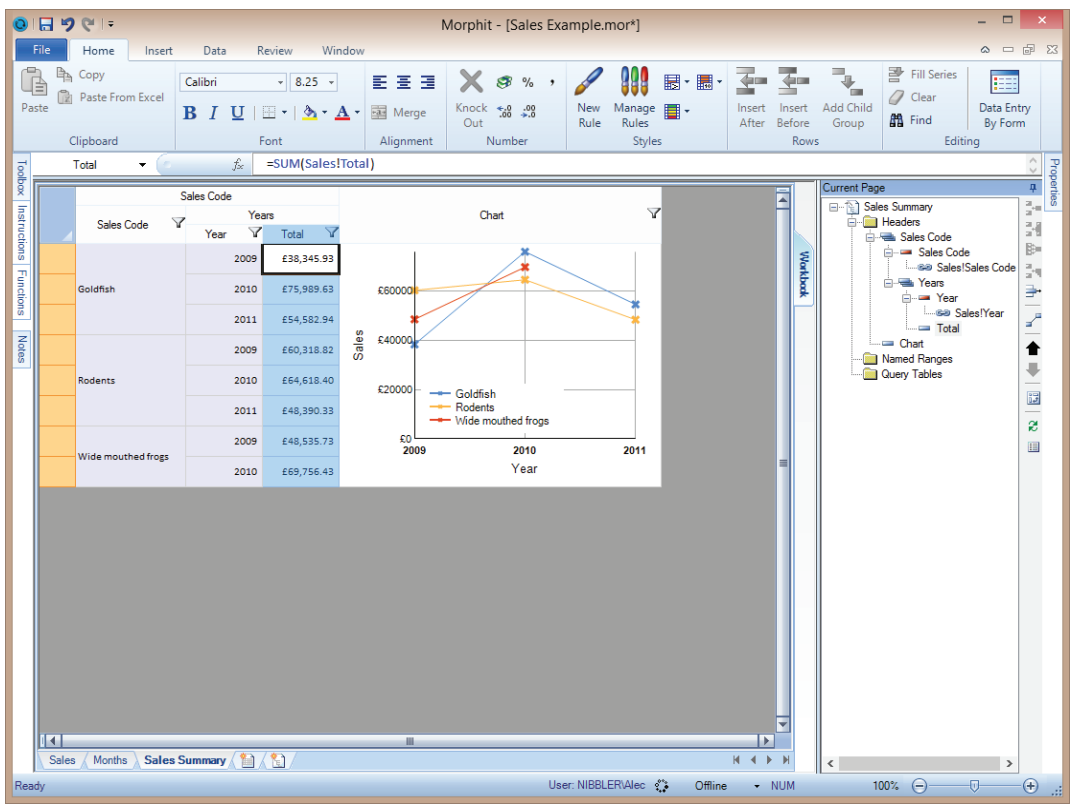

In this example we have decided that we want to take the sales data from the previous example and summarize it by sales code and year, visualizing the results on a line graph.

This *table* is constructed entirely from data sourced from the previous sales *table*. First, we borrowed in the 'Sales Code' *field*. This brings across all unique values in the 'Sales

Code' *field* in the Sales *table*. Next, we borrowed in the 'Year' *field*. This brings in all years where there are sales data matching the sales code, for example in 2011 no wide-mouthed frogs were sold, and so that year is not present.

The 'Total' *field* is then calculated as in the previous example. The formula is almost identical, however this time it references the *field* value explicitly from the 'Sales' *table* – '=SUM(Sales!Total)'. The 'Total' formula only makes use of those rows from the Sales *table* where both the 'Sales Code' and 'Year' values match reducing the scope of the formula only to related data.

If a new sales code is added to the Sales *table*, this *table* will automatically update adding a new row (and a new line in the chart). Conversely, removing all the sales data for a given sales code will remove all related data in the 'Sales Summary' *table* and chart. None of these operations require re-validation of the spreadsheet.

The technique of borrowing maximizes the use of data, removing the requirement for redundant data entry, a major source of qualitative errors [Stephen G. Powell, Kenneth R. Baker 2009].

**Linking Fields Between Tables**

Borrowing allows you to 'pull' entire *fields* across from one *table* to another. However, sometimes you might just want to lookup up values from another *table* based on the value in a particular *field*. This is illustrated by the following example with two *tables*.

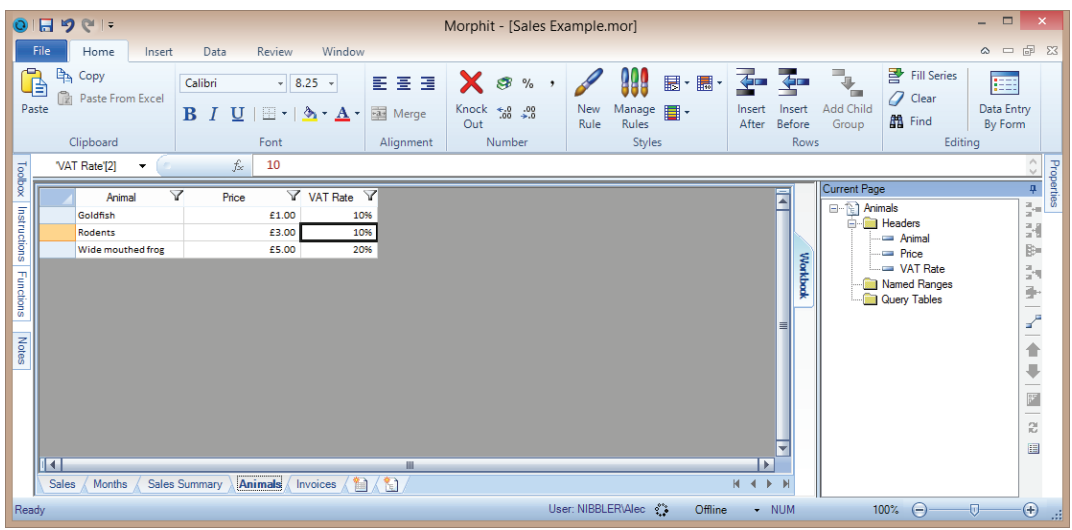

The first of these two *tables* contains a list of prices and applicable VAT rates for the animals in stock. The second *table* contains details of sales invoices. The 'Item' *field* in the 'Invoices' *table* is a drop down list containing all the animals in the 'Animals' *table*. We have then created a link between the 'Item' *field* in this *table* and the 'Animal' *field* in the 'Animals' *table*. This reduces the scope of any references to the *fields* in the 'Animals' *table* to only those rows where the 'Item' *field* value matches the 'Animal' *field* value. This is referred to as *linking through keys*. Therefore the first line of the 'Invoices' *table* identifies only one goldfish. In the 'Net' *field* the price is calculated using a simple formula '=Quantity*Animals!Price'. The 'VAT' *field* is also calculated using a similar process referencing the correct VAT rate for the selected animal.

A *table* may be linked to more than one *field* in another *table*, and to more than one other *table*. While in this example we have linked a *field* containing a drop down list, we could just as well have linked a simple text *field*.

Linking through the use of keys removes the issues of positional cell references. Formulae are scoped by the records matching the key, in a similar way to the scoping by groups. Linking is used to match imported data to existing data in *tables* removing the requirement to control the order of data and mitigating data structure errors [Stephen G. Powell, Kenneth R. Baker 2009].

3. **CONCLUSIONS**

The use of *tables* and *field* level formulae can significantly minimize physical area related errors, and enable rigorous formula auditing. The use of hierarchical *tables* and techniques such as *linking* and *borrowing* facilitate an object-orientated approach to spreadsheet design. Data can be modeled rather than just processed as happens with traditional spreadsheets. The class of object is defined by a *table*, *fields* represent the properties of the class and each row represents an object or instance of the class. Relationships between objects can be described that access data without introducing de-normalization (replicating data redundantly across *tables*). Taking this modeling approach delivers three major benefits:

1. Improved comprehension and readability.
2. Increased re-use.
3. Reduced errors.

Table-based spreadsheets are far more re-usable than traditional spreadsheets, operating under a wide variety of conditions without resorting to complex redundancy in order to incorporate all potential variations.

The robustness and inherent flexibility of spreadsheets written in this technology has already been recognized within the pharmaceutical industry, reducing the effort required to accommodate the high degree of variability within biological systems.

Whilst *tables* cannot address all spreadsheet problems, such technology should reduce the risk of error by reducing complexity and enabling rigorous auditing.

A trial version of Morphit can be downloaded from http://www.edge-ka.com.

## REFERENCES

Ayalew, Y., and Mittermeir, R. (2008), " Spreadsheet Debugging", CoRR abs/0801.4280.

Ayalew, Y., Clermont, M., and Mittermeir, R.T. (2008), " Detecting Errors in Spreadsheets", CoRR abs/0805.1740.

Butler, R.J. (2009), " The Role of Spreadsheets in the Allied Irish Bank / Allfirst Currency Trading Fraud", CoRR abs/0910.2048.

Chadwick, D. (2008a), " EuSpRIG TEAM work:Tools, Education, Audit, Management", CoRR abs/0806.0172.

Chadwick, D.R. (2008b), " Training Gamble leads to Corporate Grumble?", CoRR abs/0806.0182.

Croll, G.J. (2009), " Spreadsheets and the Financial Collapse", CoRR abs/0908.4420.

Dinmore, M. (2009), " Documenting Problem-Solving Knowledge: Proposed Annotation Design Guidelines and their Application to Spreadsheet Tools", CoRR abs/0908.1192.

Dunn, A. (2010), " Spreadsheets - the Good, the Bad and the Downright Ugly", CoRR abs/1009.5705.

Edge, T. (2013), " Morphit", Product Download Page http://www.edge-ka.com/products/morphit. 12:00pm 12/02/2013

Mittermeir, R.T., Clermont, M., and Hodnigg, K. (2008), " Protecting Spreadsheets Against Fraud", CoRR abs/0801.4268.

Panko, R.R. (2008a), " Revisiting the Panko-Halverson Taxonomy of Spreadsheet Errors", CoRR abs/0809.3613.

Panko, R.R. (2008b), " Reducing Overconfidence in Spreadsheet Development", CoRR abs/0804.0941.

Panko, R.R., and Ordway, N. (2008), " Sarbanes-Oxley: What About all the Spreadsheets?", CoRR abs/0804.0797.

Przasnyski, Z.H., Leon, L., and Seal, K.C. (2011), " In Search of a Taxonomy for Classifying Qualitative Spreadsheet Errors", CoRR abs/1111.6909.

Saadat, S. (2008), " Managing Critical Spreadsheets in a Compliant Environment", CoRR abs/0805.4211.

Stephen G. Powell, Kenneth R. Baker, B.L. (2009), " Errors in Operational Spreadsheets", Journal of Organizational and End User Computing 21, 24–36.